\documentstyle[amsfonts,aps,twocolumn]{revtex}

  2 
1

\begin{document}
\title{Uncertainty reconciles complementarity with joint measurability}

\author{P. Busch\thanks{Electronic address: p.busch@hull.ac.uk}
and C.R. Shilladay\thanks{Electronic address: c.r.shilladay@maths.hull.ac.uk}}
\address{{\small Department of Mathematics, University of Hull, Hull HU6 7RX, UK}}
\date{{\small July 2002; revised November 2002}}
\maketitle

\begin{abstract}
\noindent The fundamental principles of complementarity and uncertainty are shown to be
related to the possibility of joint unsharp measurements of pairs of noncommuting
quantum observables. A new joint measurement scheme for complementary observables is
proposed. The measured observables are represented as positive operator valued measures
(POVMs), whose intrinsic fuzziness parameters are found to satisfy an intriguing pay-off
relation reflecting the complementarity. At the same time, this relation represents an
instance of a Heisenberg uncertainty relation for measurement imprecisions. A
model-independent consideration show that this uncertainty relation is logically
connected with the joint measurability of the POVMs in question.

\vspace{12pt}

\end{abstract}

\vspace{12pt}

Ever since the inception of quantum mechanics, the principles of complementarity and
uncertainty have been fundamental issues in the debate about an adequate understanding of
this theory. Usually these principles are understood as expressions of limitations on the
preparation and measurement of quantum systems; these limitations arise from the fact
that in quantum mechanics there are many pairs of non-commuting observables.
Complementarity is the mutual exclusiveness of preparing or measuring sharp values of
certain pairs of non-commuting observables \cite{Bohr28}. The uncertainty relation gives
a more quantitative expression of complementarity: the more sharply the value of one
quantity is defined or determined, the less sharply the other quantity can be defined or
determined.

Here we will develop a somewhat more positive view of the uncertainty relation: in one
version, this relation can be understood as a pay-off inequality between the measurement
imprecisions in joint unsharp measurements of a complementary pair of observables. This
interpretation is introduced in Heisenberg's seminal paper of 1927 \cite{Heis} but at
that time and for many decades afterwards, it was not possible to formally distinguish
it from the well-known relation for variances of sharp position and momentum, and thus
it got conflated with this latter version.

It was only after the introduction of positive operator valued measures (POVMs) into
physics in the 1960s and 1970s (first by Ludwig and somewhat later by Davies, Helstrom,
Holevo, and others, for a survey, see \cite{OQP}) that a notion of joint measurement of
non-commuting observables could be formulated in the Hilbert space formalism of quantum
mechanics. This has then been used to study models of joint measurement schemes for
position and momentum, spin components and other quantities, which all led to some form
of \emph{measurement uncertainty relation}. This development is reviewed in \cite{OQP}.

In this letter we propose a new, realizable scheme of a joint measurement of
complementary spin-1/2 components (or any pair of `qubit' observables); we will find a
novel pay-off relation for measures of unsharpness associated with the observables
involved, and we will show that this form of Heisenberg \emph{measurement uncertainty
relation} is logically related to the very possibility of making a joint measurement.
This shows that the unsharpness inherent in such joint measurements is, on one hand, a
reflection of complementarity and on the other hand, a \emph{necessary and sufficient}
condition for joint measurability.

Versions of measurement uncertainty relation were found in a variety of models of joint
measurements (references to follow below), but to our knowledge, the argument given below
is the first demonstration of a logical connection: the measurement uncertainty relation
quantifies the price one has to pay for the possibility of re-establishing the classical
ideal of joint measurability in the case of non-commuting quantum observables.


The following consideration applies to all quantum systems that can be described in
terms of a 2-dimensional Hilbert space, such as spin-1/2 systems, two-level atomic
systems, or to path and interference observables in two-way interferometry. For
simplicity we use the language of spin-1/2 systems in the present section.

We will adopt the Poincar\'e sphere notation, with points on the surface represented as
Euclidean unit vectors $\bold n$ and general Pauli spin operators written as
$\sigma_{\bold n}={\bold n} \cdot \sigma=n_1\sigma_1+n_2\sigma_2+n_3\sigma_3$. We
consider any operator $\sigma_{\bold n}$ with unit vector $\bold n$ pointing to a point
in the equator as an observable complementary to $\sigma_3$.

The following experimental scheme is inspired by the work of Zhu et al \cite{Zhuetal}.
Consider a system consisting of an object $o$ and a probe $p$, both represented by
two-dimensional Hilbert spaces. The (normalized) initial state is
\begin{equation}
|\Psi_{i}\rangle=[\alpha|+\rangle_{o}+\beta
|-\rangle_{o}]\,|+\rangle_{p}=|\psi_{i}\rangle\,|+\rangle_{p},
\end{equation}
where $|\alpha|^2+|\beta|^2= 1$. The states $|+\rangle$ and
$|-\rangle$ are the eigenstates of the Pauli matrix $\sigma_3$
associated with eigenvalues 1 and -1, respectively.

Applying a suitable unitary operation \cite{Zhuetal}, the state $|\Psi_{i}\rangle$ is
transformed into an entangled state,
$$
\alpha|+\rangle_{o}|+\rangle_{p}+\beta |-\rangle_{o}|-\rangle_{p}.
$$
This state could be used to obtain unambiguous information about $\sigma_3^{(o)}$ of $o$
by measuring $\sigma_3^{(p)}$ of $p$. Application of a further unitary operation
\cite{Zhuetal}, parameterized by $\phi$, transforms this entangled state into the final
state
\begin{eqnarray}
|\Psi_{f}\rangle =&& \textstyle{\frac{1}{\surd{2}}}\big[\alpha
\left(|+\rangle_{o}-e^{-\imath\phi}
|-\rangle_{o}\right)|+\rangle_{p}\nonumber\\ &&\ \ \ +\beta
\left((e^{\imath\phi} |+\rangle_{o}
 + |-\rangle_{o}\right)|-\rangle_{p}\big].
\end{eqnarray}

It is now clear that measuring the probe observable $\sigma_3^{(p)}$ on either the
intermediate state or the final state yields information about the corresponding
observable $\sigma_3^{(o)}$ of the object: the part of the superposition containing
$|-\rangle_o|-\rangle_p$ or $|+\rangle_o|-\rangle_p$ (resp. $|-\rangle_o|+\rangle_p$ or
$|+\rangle_o|+\rangle_p$) must have evolved from $|-\rangle_o|-\rangle_p$ (resp.
$|+\rangle_o|+\rangle_p$).

We propose to go further and explore systematically the idea of exploiting the
entanglement of object and probe by \emph{jointly} measuring spin components
$\sigma_{\mathbf{o}}$ of $o$ and $\sigma_{\mathbf{p}}$ of $p$ in $|\Psi_f\rangle$,
Figure~1. It will be found that this leads to a joint measurement scheme for
complementary observables, providing information on the input state $|\psi_i\rangle$ of
the object.

We thus propose to measure an observable with the four spectral projections
${\mathrm{P}}^{o}_\pm\otimes {\mathrm{P}}^{p}_\pm$ on the object-plus-probe system in the
state $|\Psi_f\rangle$. We write ${\mathrm{P}}^{o}_{\pm}=({\mathrm{I}}\pm{\bold o}
\cdot\sigma)/2$, ${\mathrm{P}}^{p}_{\pm}=({\mathrm{I}}\pm{\bold p}\cdot\sigma)/2$, where
${{\bold o}}=
\left(\sin\theta_{o}\cos\phi_{o},\sin\theta_{o}\sin\phi_{o},\cos\theta_{o}\right)$ is a
unit vector pointing to the point with polar coordinates $(\theta_{o},\phi_{o})$.
Similarly, ${\mathbf{p}}=
\left(\sin\theta_{p}\cos\phi_{p},\sin\theta_{p}\sin\phi_{p},\cos\theta_{p}\right)$
defines the point $(\theta_{p},\phi_{p})$.

\begin{figure}
\begin{picture}(110,100)(0,0)
\put(0,50){N pairs} \put(0,40){of $o$ \& $p$}
\put(125,95){\makebox(0,0){+}} \put(125,65){\makebox(0,0){-}}
\put(125,35){\makebox(0,0){+}} \put(125,5){\makebox(0,0){-}}
\put(0,70){\framebox(20,20){$o$, $\varphi$}}
\put(0,10){\framebox(20,20){$p$, $\phi$}}
\put(45,40){\framebox(20,20){$\Psi_f$}}
\put(90,70){\framebox(20,20){$\sigma_{\mathbf{o}}$}}
\put(90,10){\framebox(20,20){$\sigma_{\mathbf{p}}$}}
\put(125,95){\circle{10}} \put(125,65){\circle{10}}
\put(125,35){\circle{10}} \put(125,5){\circle{10}}
\put(65,50){\vector(1,1){25}} \put(65,50){\vector(1,-1){25}}
\put(20,80){\vector(1,-1){25}} \put(20,20){\vector(1,1){25}}
\put(110,80){\vector(1,1){10}} \put(110,80){\vector(1,-1){10}}
\put(110,20){\vector(1,1){10}} \put(110,20){\vector(1,-1){10}}
\put(130,95){\vector(1,0){30}} \put(130,95){\vector(1,-1){30}}
\put(130,35){\vector(1,2){30}} \put(130,35){\vector(1,1){30}}
\put(130,65){\vector(1,-1){30}} \put(130,65){\vector(1,-2){30}}
\put(130,5){\vector(1,0){30}} \put(130,5){\vector(1,1){30}}
%
\put(170,95){\oval(20,10)} \put(170,95){\makebox(0,0){+ +}}
\put(170,65){\oval(20,10)} \put(170,65){\makebox(0,0){+ -}}
\put(170,35){\oval(20,10)} \put(170,35){\makebox(0,0){- +}}
\put(170,5){\oval(20,10)} \put(170,5){\makebox(0,0){- -}}
\put(180,95){\vector(1,0){10}}\put(191,90){$N_{++}$}
\put(180,65){\vector(1,0){10}}\put(191,60){$N_{+-}$}
\put(180,35){\vector(1,0){10}}\put(191,30){$N_{-+}$}
\put(180,5){\vector(1,0){10}}\put(191,0){$N_{--}$}
\end{picture}\\
\caption{N pairs of object $o$ and probe $p$ systems are entangled into the state
$\psi_{f}$ (eq. 2). Measurements of spin are made separately on $o$ and $p$ and pairs of
outcomes counted. In an ideal experiment the frequencies $N_{++}$, $N_{+-}$, $N_{-+}$,
$N_{--}$ will add up to $N$.}
\end{figure}
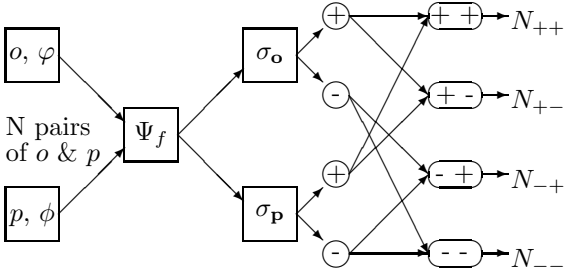

Our aim is to write the probabilities for the pair outcomes in terms of the input state
$|\psi_{i}\rangle$. This will give a POVM with four positive operators (referred to as
\emph{effects} \cite{effects}) $\mathrm{F}_{\pm,\pm}$ adding up to the unity,
$\mathrm{I}$. For example, the effect $\mathrm{F_{++}}$ will be uniquely determined by
the condition
\begin{eqnarray}\label{condition}
\langle\psi_i|{\mathrm{F}}_{++}\psi_i\rangle=
\textstyle\langle\Psi_f|{\mathrm{P}}^o_{+}\otimes{\mathrm{P}}^p_{+}\Psi_f\rangle,
\end{eqnarray}
which is to hold for all $|\psi_i\rangle$. After a long run of counting it can be
assumed that the frequency, for example, of $N_{++}\approx
N\langle\psi_i|{\mathrm{F}}_{++}\psi_i\rangle $ (Figure 1). The condition of eq.
\ref{condition} defines the object observable measured on the input state
$|\psi_i\rangle$ by measuring the observable with the four spectral projections
${\mathrm{P}}^{o}_{\pm}\otimes{\mathrm{P}}^p_{\pm}$ on the output state
$|\Psi_f\rangle$. One finds, \cite{Shil}
\begin{eqnarray}\label{4povm}
{\mathrm{F}}_{++}&=&{\textstyle\frac{1}{4}}\left((1+\mathrm{A}\mathrm{B})\mathrm{I}
+N_{1}\sigma_{1} + N_{2}\sigma_{2}
+(\mathrm{A}+\mathrm{B})\sigma_{3}\right),\nonumber\\
\mathrm{F}_{+-}&=&\textstyle{\frac{1}{4}}\left((1-\mathrm{A}\mathrm{B})\mathrm{I}-
N_{1}\sigma_{1} -
N_{2}\sigma_{2}+(\mathrm{A}-\mathrm{B})\sigma_{3}\right),\nonumber\\
\mathrm{F_{-+}}&=&\textstyle{\frac{1}{4}}\left((1-\mathrm{A}\mathrm{B})\mathrm{I}-
N_{1}\sigma_{1} -
N_{2}\sigma_{2}-(\mathrm{A}-\mathrm{B})\sigma_{3}\right),\\
\mathrm{F_{--}}&=&\textstyle{\frac{1}{4}}\left((1+\mathrm{A}\mathrm{B})\mathrm{I}+
N_{1}\sigma_{1} + N_{2}\sigma_{2} -(\mathrm{A}+\mathrm{B})\sigma_{3}\right).\nonumber
\end{eqnarray}
Here, ${\mathrm{A}}=-\sin\theta_{o}\cos(\phi_{o}+\phi)$, ${\mathrm{B}}=\cos\theta_{p}$;
$\mathrm{N_{1}}=\mathrm{W}\mathrm{Y}+\mathrm{X}\mathrm{Z}$,
$\mathrm{N_{2}}=-\mathrm{X}\mathrm{Y}+\mathrm{W}\mathrm{Z}$, with
${\mathrm{W}}=\cos(\phi-\phi_{p})$, ${\mathrm{X}}=\sin(\phi-\phi_{p})$,
${\mathrm{Y}}=\cos\theta_{o}\sin\theta_{p}$,
${\mathrm{Z}}=\sin\theta_{o}\sin(\phi_{o}+\phi)\sin\theta_{p}$.

These four effects can be grouped into two pairs in three different ways, and summing
the two elements of each pair gives a new pair of effects adding up to $\mathrm{I}$. In
this way one obtains three marginal POVMs:
\begin{eqnarray}\label{margs}
\mathrm{F}_{+}^{(1)}&=&\mathrm{F}_{++} + \mathrm{F}_{+-}=
\textstyle{\frac{1}{2}}(\mathrm{I}+\mathrm{A}\sigma_3),\nonumber\\
\mathrm{F}_{-}^{(1)}&=&\mathrm{F}_{-+} + \mathrm{F}_{--} =
\textstyle{\frac{1}{2}}(\mathrm{I}-\mathrm{A}\sigma_3);\nonumber\\
\mathrm{F}_{+}^{(2)}&=&\mathrm{F}_{++} + \mathrm{F}_{-+}=
\textstyle{\frac{1}{2}}(\mathrm{I}+\mathrm{B}\sigma_3),\nonumber\\
\mathrm{F}_{-}^{(2)}&=&\mathrm{F}_{+-} +\mathrm{F}_{--}=
\textstyle{\frac{1}{2}}(\mathrm{I}-\mathrm{B}\sigma_3);\\
\mathrm{F}_{+}^{(3)}&=&\mathrm{F}_{++} + \mathrm{F}_{--}=
{\textstyle{\frac
12}}\left((1+\mathrm{A}\mathrm{B}){\mathrm{I}}+\mathrm{N_{1}}\sigma_1+
\mathrm{N_{2}}\sigma_2\right),\nonumber\\
\mathrm{F}_{-}^{(3)}&=&\mathrm{F}_{+-} + \mathrm{F}_{-+}={\textstyle{\frac
12}}\left((1-\mathrm{A}\mathrm{B}){\mathrm{I}}-\mathrm{N_{1}}\sigma_1-
\mathrm{N_{2}}\sigma_2\right).\nonumber
\end{eqnarray}
The first two marginal POVMs are smeared versions of the sharp observable
$\sigma_3^{(o)}$ since their effects are combinations of the spectral projections of
$\sigma_3^{(o)}$. The third POVM is a smeared version of an observable complementary to
$\sigma_3^{(o)}$; its spectral decomposition is in terms of the spectral measure of
$\sigma_{\mathbf{n}}^{(o)}$, with ${\mathbf{n}}=\mathbf{N}/|\mathbf{N}|$,
${\mathbf{N}}=({\mathrm{N_{1}}},{\mathrm{N_{2}}},0)$, and $\mathbf{n}$ perpendicular to
the vector $(0,0,1)$ associated with $\sigma_3^{(o)}$.

Using this measurement scheme as illustrated in Figure 1 it is possible to compare the
marginal probabilities with the probabilities obtained in a sharp measurement. For
example,
\begin{eqnarray}\label{prob1}
\langle\psi_i|{\mathrm{F}}_{+}^{(1)}\psi_i\rangle&=&
\textstyle\frac{1}{2}\left(1+\mathrm{A}\langle\psi_i|\sigma_3^{(o)}\psi_i\rangle\right)\nonumber\\
&=&
\textstyle\frac{1}{2}\left(1+\mathrm{A}\langle\psi_i|\mathrm{P}^{\sigma_3}_+\psi_i\rangle-
\mathrm{A}\langle\psi_i|\mathrm{P}^{\sigma_3}_-\psi_i\rangle\right)\nonumber\\
&=&\mathrm{A}\langle\psi_i|\mathrm{P}^{\sigma_3}_+\psi_i\rangle+
\textstyle\frac{1}{2}(1-\mathrm{A})
\end{eqnarray}
where $\mathrm{P}^{\sigma_3}_+,\
\mathrm{P}^{\sigma_3}_-={\mathrm{I}}-\mathrm{P}^{\sigma_3}_+$ are the spectral
projections of $\sigma_3^{(o)}$. The number
$\langle\psi_i|{\mathrm{F}}_{+}^{(1)}\psi_i\rangle
=\langle\psi_i|{\mathrm{F}}_{++}\psi_i\rangle
+\langle\psi_i|{\mathrm{F}}_{+-}\psi_i\rangle\approx(N_{++}+N_{+-})/N$ is the
probability of a + outcome in an unsharp measurement of $\sigma_3^{(o)}$, represented
here by the marginal POVM $(F_+^{(1)},F_-^{(1)})$. Eq.~\ref{prob1}\ can be solved for
the probability $\langle\psi_i| \mathrm{P}^{\sigma_3}_+\psi_i\rangle$ of a + outcome in
a sharp measurement of $\sigma_3^{(o)}$. In this way the frequencies $N_{++}$ and
$N_{+-}$ obtained in our scheme can be used to reconstruct this sharp measurement
probability.

Similarly we calculate:
\begin{eqnarray}
\langle\psi_i|{\mathrm{F}}_{+}^{(3)}\psi_i\rangle&=&\textstyle\frac 12(1+AB)+
\textstyle\frac 12\langle\psi_i|{\mathbf{N}}\cdot\sigma\,\psi_i\rangle\nonumber\\
&=&\textstyle\frac 12(1+AB-|{\mathbf{N}}|)+ |{\mathbf{N}}|\langle\psi_i
|\mathrm{P}_+^{\mathbf{n}} \psi_i\rangle.
\end{eqnarray}
Here we have used the fact that the spectral projections of $\sigma_{\mathbf{n}}^{(o)}$
satisfy $\mathrm{P}_+^{\mathbf{n}}+\mathrm{P}_-^{\mathbf{n}}=\mathrm{I}$. We see again
that estimation of the third marginal probability (by way of collecting the counts
$N_{++}+N_{--}$) allows one to reconstruct the probability for a + outcome of a sharp
measurement of $\sigma_{\mathbf{n}}^{(o)}$, where this observable is complementary to
$\sigma_3^{(o)}$. We thus obtain a simultaneous determination of the probabilities for
two complementary observables from the statistics of a single experiment. Similar
calculations can be made for a general mixed state. We thus see that our joint
measurement scheme provides information which is equivalent to the knowledge of two of
the three parameter values characterizing a general state.

We can see Bohr's principle of complementarity at work in the present joint measurement
scheme. Consider the case in which ${\mathrm{F}}^{(3)}_\pm$ becomes a projection, hence
a sharp observable complementary to $\sigma_3^{(o)}$; that is:
$\mathrm{N_{1}}^{2}+\mathrm{N_{2}}^{2}$=1. In this case,
\begin{eqnarray*}
{\mathrm{N_{1}}}^{2}+{\mathrm{N_{2}}}^{2}=
\sin^{2}\theta_{p}[\cos^{2}\theta_{o}+\sin^{2}\theta_{o} \sin^{2}(\phi_{o}+\phi)]=1,
\end{eqnarray*}
hence $\sin^{2}\theta_{p}$=1 and $\sin^{2}(\phi_{o}+\phi)$=1 or $\sin^{2}\theta_{p}$=1
and $\cos^{2}\theta_{o}$=1. Either case implies that
$\mathrm{A}^{2}=\sin^{2}\theta_{o}\cos^{2}(\phi_{o}+\phi)$=0 and
$\mathrm{B}^{2}=\cos^{2}\theta_{p}$=0 i.e. $\mathrm{A}\mathrm{B}$=0. So, demanding that
${\mathrm{F}}_{\pm}^{(3)}$ is a sharp observable, means that there is no information
about $\sigma_3^{(o)}$, as the first two marginal POVMs become trivial:
$\mathrm{F}_{\pm}^{(1)}=\mathrm{F}_{\pm}^{(2)}$=$\mathrm{I}/2$.

Conversely, if we require that $\mathrm{F}_{\pm}^{(2)}$ is a sharp observable, then
$|\mathrm{B}|$=$|\cos\theta_{p}|=1$ and $\mathrm{N_{1}}$=$\mathrm{N_{2}}=0$, giving
$\mathrm{F}^{(3)}_{\pm}=(1\pm|\mathrm{A}|)\mathrm{I}/2$. Similarly, if
$\mathrm{F}_{\pm}^{(1)}$ is a sharp observable, then
$\mathrm{F}^{(3)}_{\pm}=(1\pm|\mathrm{B}|)\mathrm{I}/2$. Hence we find for this joint
measurement scheme: if one of a pair of complementary marginal observables is sharp the
other is trivialized.

It is possible to go further and give a quantitative relationship
expressing this form of \emph{measurement complementarity}. A
natural measure for the quality, or \emph{contrast}, of the
smeared marginal observables given above is the difference between
the largest and smallest eigenvalues of each of the effects
$\mathrm{F}^{(j)}_\pm$. This quantity can be determined as the
greatest possible probability for the associated outcome, minus
the smallest possible probability. In the case of a projection,
that number is 1-0=1. Thus, the closer the contrast of one of the
above marginal POVMs is to 1, the closer its effects are to
projections.

The eigenvalues of both effects of the first marginal POVM, $\mathrm{F}^{(1)}_{\pm}$, are
$(1\pm\mathrm{A})/2$ and those of the second marginal, $\mathrm{F}^{(2)}_{\pm}$, are
$\left(1\pm\mathrm{B}\right)/2$. The effect $\mathrm{F}^{(3)}_{+}$ of the third marginal
has the eigenvalues $(1+\mathrm{A}\mathrm{B})/2 \pm  \allowbreak
{(\mathrm{N_{1}}^{2}+\mathrm{N_{2}}^{2})^{1/2}}/2$. Those of the other effect,
$\mathrm{F}^{(3)}_{-}$, of the third marginal are $(1-\mathrm{A}\mathrm{B})/2
\pm{(\mathrm{N_{1}}^{2}+\mathrm{N_{2}}^{2})^{1/2}}/2$.

We thus obtain for the contrast of each of the marginals $\mathrm{F}^{(1)}$,
$\mathrm{F}^{(2)}$, and $\mathrm{F}^{(3)}$ the values $\mathrm{C}_1=|\mathrm{A}|$,
$\mathrm{C}_2=|\mathrm{B}|$, and
$\mathrm{C}_3={(\mathrm{N_{1}}^2+\mathrm{N_{2}}^2)^{1/2}}$, respectively. After some
manipulation it can be seen that,
\begin{eqnarray}
\mathrm{N_{1}}^{2}+\mathrm{N_{2}}^{2}&=&\left(1-\cos^{2}\theta_{p}\right)
\left(1-\sin^2\theta_{o}\cos^{2}(\phi_{o})\right)\nonumber\\
&=&(1-\mathrm{A}^{2})(1-\mathrm{B}^{2}),
\end{eqnarray}
that is,
\begin{equation}\label{payoff}
\mathrm{C}_3^2=\left(1-\mathrm{C}_1^2\right)\left(1-\mathrm{C}_2^2\right).
\end{equation}

Here eq.~\ref{payoff} is a state independent relationship between two unsharp
$\sigma_3^{(o)}$ observables and the third marginal observable. It says that there is a
pay-off: the better the contrast of the latter observables, the the worse the contrast
of the two $\sigma_3^{(o)}$  marginals must be; and vice versa.

We can take $1-\mathrm{C}^2$ as a measure of the intrinsic
unsharpness, or fuzziness of the POVM. This becomes evident if one
calculates the variance of, say, the POVM $\mathrm{F}^{(1)}$,
considered as a smeared version of the observable $\sigma_3$ which
has values $\pm 1$. Using
${\mathrm{Var}}_\rho({\mathrm{F}})=\int(t-\overline{t})^2\,d\langle
{\mathrm{F}}_t\rangle_\rho$, $\overline{t}=\int t\,d\langle
{\mathrm{F}}_t\rangle_\rho$, and putting
$\rho=({\mathrm{I}}+{\mathbf{r}}\cdot\sigma)/2$, we obtain
$\overline{t}=\langle {\mathrm{F}}^{(1)}_+\rangle_\rho-\langle
{\mathrm{F}}^{(1)}_-\rangle_\rho={\mathrm{A}}r_3$, and
\begin{eqnarray}\label{variance}
{\mathrm{Var}}_\rho({\mathrm{F}}^{(1)})&=&1-{\mathrm{A}}^2r_3^2\nonumber\\
&=&{\mathrm{Var}}_\rho(\sigma_3)+
(1-{\mathrm{A}}^2)(1-{\mathrm{Var}}_\rho(\sigma_3)).
\end{eqnarray}
Here we have used the relation ${\mathrm{Var}}_\rho(\sigma_3)=1-r_3^2$. The minimum of
the variance in eq.~\ref{variance} over all $\rho$ is obtained at eigenstates of
$\sigma_3$, where one obtains the value $1-{\mathrm{A}}^2$ (as $r_3=\pm 1$). This
minimal spread of outcomes reflects the intrinsic fuzziness of the measurement, that is,
the uncertainty about the actual value prior to measurement.

Equation \ref{payoff} implies the following pay-off relations for the complementary
\emph{pairs} of marginals:
\begin{equation}\label{pairpayoff}
\left(\mathrm{N_{1}}^{2}+\mathrm{N_{2}}^{2}\right)+ \mathrm{A}^{2}\leq 1,\ \ \
\left(\mathrm{N_{1}}^{2}+\mathrm{N_{2}}^{2}\right)+\mathrm{B}^{2}\leq 1.
\end{equation}
Which in terms of the contrasts are,
\begin{equation}\label{pairpayoff2}
\mathrm{C}_{3}^{2}\leq 1-\mathrm{C}_{1}^{2},\ \ \
\mathrm{C}_{3}^{2}\leq 1-\mathrm{C}_{2}^{2},
\end{equation}
or, introducing the \emph{unsharpness}
${\mathrm{U}}_i=1-{\mathrm{C}}_i^2$,
\begin{equation}\label{HUR}
{\mathrm{U}}_1+{\mathrm{U}}_3\ge 1,\ \ \
{\mathrm{U}}_2+{\mathrm{U}}_3\ge 1.
\end{equation}

It appears as if these relations  are consequences of the fact that the observables
$\mathrm{F}^{(1)}$, $\mathrm{F}^{(2)}$ and $\mathrm{F}^{(3)}$ are jointly measurable.
Similar results were obtained in different joint measurement models, and with other
measures of unsharpness, in \cite{Bu85,MdeM,Luis}. In the present case, we can go one
step further and show that conditions of the form of eqs.~\ref{pairpayoff2} and
\ref{HUR} are in fact \emph{necessary and sufficient} conditions for the joint unsharp
measurability of complementary (qubit/spin-1/2) observables.

In \cite{Bu86}, one of the authors formulated necessary and sufficient conditions for
unsharp spin-1/2 observables to be jointly measurable (i.e. their effects occur in the
range of a common POVM). A pair of two valued POVMs,
$\big\{{\mathrm{F}}^{(1)}_{\pm}=\left({\mathrm{I}}\pm{\bold a}
\cdot\sigma\right)\big/2\}$ and $\big\{{\mathrm{F}^{(3)}_{\pm}}=\left({\mathrm{I}}
\pm{\bold b}\cdot\sigma\right)\big/2\}$ is jointly measurable exactly when $|{\bold
a}+{\bold b}|/2 +|{\bold a}-{\bold b}|/2\leq1$. If $\bold a$ and $\bold b$ are
perpendicular, the two POVMs represent unsharp versions of complementary observables. In
this case the coexistence condition assumes the form,
\begin{equation}
|{\bold a}|^{2}+|{\bold b}|^{2}\leq 1.
\end{equation}
This is indeed equivalent to the pay-off relationships  of eq. \ref{pairpayoff} deduced
from the model where ${\bold a}=(0,0,-\mathrm{A})$ or ${\bold a}=(0,0,\mathrm{B})$, and
${\bold b}= (\mathrm{N_{1}},\mathrm{N_{2}},0)$.

If we put $\mathrm{B}=0$ into the model, the pay-off relation reduces to
$(\mathrm{N_{1}}^{2}+\mathrm{N_{2}}^{2})+ \mathrm{A}^{2}= 1$. Hence this  relation,
taken as an expression of complementarity and deduced from the  model, is equivalent to
the joint measurability condition in the special case where $\mathrm{B}=0$ (or similarly
where $\mathrm{A}=0$).

We leave it as an open question as to what the joint measurability condition looks like
for a pair of POVMs ${\mathrm{F}}^{(1)}_\pm=\left({\mathrm{I}}\pm A\sigma_3\right)/2$,
${\mathrm{F}}^{(3)}_\pm=\left((1-L){\mathrm{I}}
\pm({\mathrm{N_{1}}},{\mathrm{N_{2}}},0)\cdot\sigma\right)/2$. It would also be of
interest to establish a criterion for the joint measurability of a triple of POVMs of
the form of eq. \ref{margs}.


The joint measurement scheme proposed here can be realized as an extension of the
experiment of Zhu et al \cite{Zhuetal}. This experiment uses nuclear magnetic resonance
techniques to entangle the spin states of the $^{13}C$ nucleus with those of the $^{1}H$
nucleus in a chloroform molecule, $^{13}CHCl_{3}$. The  $^{13}C$ nucleus is taken to be
the object system ($o$), while the $^{1}H$ nucleus serves as the probe ($p$). The
unitary operations used to transform the initial state $|\Psi_i\rangle$ into the final
state $|\Psi_f\rangle$ are realized in this experiment by application of electromagnetic
pulse sequences with appropriate frequencies or field gradients.

Zhu et al claim without proof to have obtained perfect population and coherence data from
the statistics of a single run of an experiment, and thereby they seem to suggest that
complementarity is not unconditionally valid but only subject to some qualifications. Our
analysis shows that the joint unsharp measurability of complementary observables is
consistent with complementarity.

Moreover, we have seen that the statistics obtained allow one to reconstruct the
probabilities for sharp measurements of two spin components of $o$, namely, $\sigma_3$
and $\sigma_{\mathbf{n}}$. This means that the present experimental scheme constitutes an
effective state determination procedure in that two out of three parameters
characterizing any quantum state can be measured. In particular, this confirms that the
population and coherence data (as defined in\cite{Zhuetal}) can indeed be recovered from
such a joint measurement scheme.

A complementarity, or duality relation of the form $V^2+D^2\le 1$ has been deduced in
\cite{JaShiVai,Eng} in the context of two-path interferometry as a pay-off between a
measure $D$ of path distinguishability and the visibility $V$ of interference fringes.
These measures are characteristics of the quantum state $\rho$ and can be determined
separately in measurements of sharp path and interference observables.

The inequality $V^2+D^2\le 1$ bears resemblance in form to eq.~\ref{pairpayoff2} but
their significance is fundamentally different: the quantities $V,D$ refer to the object
state and to mutually exclusive measurement of sharp path and interference observables; a
generalization to arbitrary pairs of unsharp path and interference observables is
straightforward \cite{Bjorketal}. By contrast, eq.~\ref{pairpayoff2} describes a
state-independent relation between specific pairs of unsharp path and interference
observables, namely those which are jointly measurable. The former relation is an
expression of complementarity for preparations while the latter reflects measurement
complementarity -- and at the same time enables joint measurability.

To conclude: we have shown that for complementary pairs of `qubit' observables, the very
possibility of making joint unsharp measurements is logically connected with the
measurement version of Heisenberg's uncertainty principle. This connection is illustrated
in a new realizable joint measurement scheme, the analysis of which gives rise to a novel
pay-off relation for the contrasts of the measured unsharp observables. The contrast
measures thus found enable an operational interpretation of a joint measurability
condition found earlier by one of the authors. Finally, the measurement scheme discussed
here provides an illustration of the way in which the use of entanglement (between the
object and a probe) can lead to more powerful state determination procedures.


\begin{references}
\bibitem{effects} This term was coined by the Marburg school of G. Ludwig. For original
references, cf. Ref. \cite{OQP}.
\end{references}
\end{document}